%% file: main.tex
\documentclass[letterpaper,conference]{IEEEtran}
\IEEEoverridecommandlockouts
\input{header.tex}

\begin{document}
\title{An Achievable Rate-Distortion Region for\\
Joint State and Message Communication over Multiple Access Channels
\thanks{The authors were supported in part by the
    German Federal Ministry of Education and Research (BMBF)
    in the programme “Souver\"an. Digital. Vernetzt.”
    within the research hub 6G-life under Grant 16KISK002,
    and also by the Bavarian Ministry of Economic Affairs,
    Regional Development and Energy within the project 6G Future Lab Bavaria.
    U. M{\"o}nich and H. Boche were also supported by the BMBF within the project "Post Shannon Communication - NewCom" under Grant 16KIS1003K.
    }
}

\IEEEoverridecommandlockouts\IEEEpubid{\makebox[\columnwidth]{ 978-1- 979-8-3503-1090-0/23/\$31.00~\copyright~2023 European Union\hfill} \hspace{\columnsep}\makebox[\columnwidth]{ }}

\makeatletter
\def\ps@IEEEtitlepagestyle{%
  \def\@oddfoot{\mycopyrightnotice}%
  \def\@oddhead{\hbox{}\@IEEEheaderstyle\leftmark\hfil\thepage}\relax
  \def\@evenhead{\@IEEEheaderstyle\thepage\hfil\leftmark\hbox{}}\relax
  \def\@evenfoot{}%
}
\def\mycopyrightnotice{%
  \begin{minipage}{\textwidth}
  \centering \scriptsize
  Copyright~\copyright~20XX IEEE.  Personal use of this material is permitted.  Permission from IEEE must be obtained for all other uses, in any current or future media, including reprinting/republishing this material for advertising or promotional purposes, creating new collective works, for resale or redistribution to servers or lists, or reuse of any copyrighted component of this work in other works.
  \end{minipage}
}
\makeatother

\author{\IEEEauthorblockN{Xinyang Li\IEEEauthorrefmark{1}\IEEEauthorrefmark{2},
Vlad C. Andrei\IEEEauthorrefmark{1}\IEEEauthorrefmark{3}, Ullrich J. M{\"o}nich\IEEEauthorrefmark{1}\IEEEauthorrefmark{4} and
Holger Boche\IEEEauthorrefmark{1}\IEEEauthorrefmark{5}}
\IEEEauthorblockA{\IEEEauthorrefmark{1}Chair of Theoretical Information Technology, Technical University of Munich, Munich, Germany\\
\IEEEauthorrefmark{1}BMBF Research Hub 6G-life,
\IEEEauthorrefmark{5}Munich Center for Quantum Science and Technology,
\IEEEauthorrefmark{5}Munich Quantum Valley\\
Email: \IEEEauthorrefmark{2}xinyang.li@tum.de,
\IEEEauthorrefmark{3}vlad.andrei@tum.de,
\IEEEauthorrefmark{4}moenich@tum.de,
\IEEEauthorrefmark{5}boche@tum.de}
}

\maketitle

\allowdisplaybreaks
\glsdisablehyper

\begin{abstract}
This paper derives an achievable \ac{rd} region for the \ac{sddmmac}, where the generalized feedback and causal side information are present at encoders, and the decoder performs the joint task of message decoding and state estimation. 
The Markov coding and backward-forward two-stage decoding schemes are adopted in the proof. This scenario is shown to be capable of modeling various \ac{isac} applications, including the monostatic-uplink system and multi-modal sensor networks, which are then studied as examples.
\end{abstract}


\glsresetall

\section{Introduction}\label{sec:intro}
\input{sections/intro.tex}

\section{Channel Model}
\input{sections/model}

\section{Main Results}
\input{sections/results}

\section{Examples and Applications}
\input{sections/app}

\section{Proof of Achievability}\label{sec:proof}
\input{sections/proof}

\section{Conclusion}
\input{sections/conclusion}

\bibliographystyle{IEEEtran}
\bibliography{IEEEabrv,mybib.bib}
\end{document}

%% file: header.tex
\usepackage{amsmath, bm, amssymb, algpseudocode, algorithm, mathtools, svg,ifluatex, pgfplots, amsthm, caption, subcaption, tikz}
\usepackage{cite, bbm, accents}
\usepackage{graphicx}
\usepackage{textcomp}
\usepackage{xcolor}
\usepackage{stfloats,cuted}
\usepackage[acronym,shortcuts]{glossaries}
\usepackage[letterpaper, bottom=1.05in, top=0.75in, left=0.63in, right=0.63in]{geometry}
\usetikzlibrary{graphs, positioning, quotes, shapes.geometric, arrows}

\pgfplotsset{compat=1.15}

\newacronym{isac}{ISAC}{integrated sensing and communication}
\newacronym{iid}{i.i.d}{independently and identically distributed}
\newacronym{sit}{SI-T}{side information at transmitter}
\newacronym{sir}{SI-R}{side information at receiver}
\newacronym{mac}{MAC}{multiple access channel}
\newacronym{sddmmac}{SD-DMMAC}{state-dependent discrete memoryless multiple access channel}
\newacronym{cd}{C-D}{capacity-distortion}
\newacronym{bs}{BS}{base station}
\newacronym{ue}{UE}{user equipment}
\newacronym{ap}{AP}{access point}
\newacronym{bc}{BC}{broadcast}
\newacronym{rd}{R-D}{rate-distortion}

\newcommand*{\brrr}[1]{\left \{ #1 \right \}}

\newcommand{\tR}{\tilde{R}}
\newcommand{\typset}{\mathcal{T}^{(n)}}

\newcommand*{\indf}[1]{\mathbbm{1}\brrr{#1}}

\newcommand{\expcs}[2]{\mathbb{E}_{#1}[#2]}

\newcommand{\calP}{\mathcal{P}}
\newcommand{\calR}{\mathcal{R}}
\newcommand{\calC}{\mathcal{C}}
\newcommand{\hS}{\hat{S}}

\DeclareMathAccent{\bar}{\mathalpha}{operators}{"16}

\DeclareMathOperator*{\argmin}{arg\,min}

\newtheorem{theorem}{Theorem}
\newtheorem{lemma}{Lemma}

\theoremstyle{definition}

\newtheorem{remark}{Remark}

\tikzset{%
    smallblock/.style={draw, fill=white, minimum height=1em, minimum width=1em},
    block-common/.style={draw, fill=white, minimum height=1.5em, minimum width=4em},
    block/.style={rectangle, block-common, align=center},
    txtblock/.style={block, align=center, minimum height=4em},
    bigblock/.style={block, minimum height=8em},
    txtbigblock/.style={bigblock, align=center},
    input/.style={inner sep=1pt},       
    output/.style={inner sep=1pt},      
    sum/.style = {draw, fill=white, circle, minimum size=1.1em, inner sep=0pt,
      font={\small$+$}},
    prod/.style = {draw, fill=white, circle, minimum size=1.1em, inner sep=0pt,
      font={\normalsize$\times$}},
    pinstyle/.style = {pin edge={to-,thin,black}}
}

%% file: sections/intro.tex
The study on \ac{cd} trade-offs bridges the gap between estimation theory and information theory. This framework provides a methodology to investigate the performance limits of the communication systems, having the goal to not only convey messages reliably but also estimate the channel state accurately\cite{sutivong2003channel,kim2008state,zhang2011joint,choudhuri2013causal,bross2017rate}. 
With the growing focus on \ac{isac} technology, which is envisioned to be standardized in the next generation wireless networks~\cite{liu2022integrated, schwenteck20236g}, it becomes increasingly important to analyze its system performance from the perspective of \ac{cd} trade-offs\cite{ahmadipour2022information,li2024analysis,ahmadipour2022coding}. 
The authors in \cite{ahmadipour2022information} take the first step to analyze such trade-offs for a monostatic-downlink \ac{isac} system, by deriving the optimal \ac{cd} function for the point-to-point case as well as an outer and inner bound for the \ac{cd} region of \ac{bc} channels.
\cite{li2024analysis} extends the point-to-point and \ac{bc} scenarios by adding side information at both transmitter and receiver, which can be interpreted in different ways to model more generalized systems. 
Following the same monostatic-downlink setting as \cite{ahmadipour2022information}, an achievable \ac{rd} region for the \ac{mac} is developed in \cite{ahmadipour2022coding} by additionally encoding the state information contained in the generalized feedback into transmit signals, in order to improve the estimation task at the other encoder. 

In this paper, we build a \ac{mac} model based on\cite{li2024analysis}, 
where a decoder tries to decode the messages from two encoders and simultaneously estimate the channel state. Different from\cite{ahmadipour2022coding}, we assume that encoders have access to partial state information, and the estimation task is performed at the decoder side.
This problem is challenging since even without distortion constraints, the capacity regions of special cases with either generalized feedback or side information are still not determined.
Nonetheless, an inner bound of the \ac{cd} region is derived for the proposed model. The adopted coding scheme enables partial collaboration between encoders through the feedback links~\cite{willems1982informationtheoretical,carleial1982multiple,kramer2008topics}, and on the other hand designs two-stage state descriptions
to enhance both communication and estimation performance~\cite{li2024analysis,lapidoth2012multiple}. 
This channel model captures multiple \ac{isac} systems, particularly the monostatic-uplink \ac{isac}, in which case the proposed \ac{rd} region is shown to be the \ac{cd} region.

\textit{Notations:} This paper follows the same notations as~\cite{li2024analysis}. For an arbitrary joint distribution $P_{SW}$ and a distortion measurement $d(S,\hS)$, the optimal estimator\cite[Lemma 5]{li2024analysis} of $S$ given the observation $W=w$ is given by 
\begin{equation}\label{eq:opt_est}
    h^*(w) = \argmin_{\hat{s}} \expcs{}{d(S,\hat{s}) | W=w}.
\end{equation}

%% file: sections/model.tex
\begin{figure}[!t]
    \centering
    \input{sections/figures/channel_model.tikz}
    \caption{The \acs{sddmmac} with generalized feedback and side information.}
    \label{fig:channelmodel}
\end{figure}
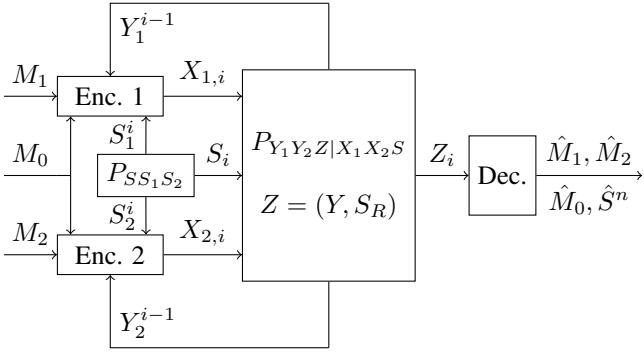

The \ac{sddmmac} studied in this paper is illustrated in Fig.~\ref{fig:channelmodel}.
The common message $M_0 \in [2^{nR_0}]$ and the individual private messages $M_1\in [2^{nR_1}]$, $M_2 \in [2^{nR_2}]$ are encoded into the transmit signals $X_{1,i}$ and $X_{2,i}$ at time step $i\in [n]$, respectively. The channel is characterized by the transition distribution $P_{Y_1Y_2Z|X_1X_2S}$ independent of $i$, and puts out the generalized feedback $Y_{1,i-1}$, $Y_{2,i-1}$ to both encoders as well as $Z_i$ to the decoder. Upon receiving $Z_i$, the decoder jointly recovers the three messages and channel state.
We indicate explicitly $Z=(Y,S_R)$ with the normal channel output $Y$ and \ac{sir} $S_R$ to emphasize the fact that \ac{sir} can be considered  as another channel output\cite{ahmadipour2022coding,li2024analysis}.
The channel state $S_i$ at time step $i$ is generated \ac{iid} according to $P_S$, and both encoders additionally have access to their respective \ac{sit} $S_1^i$ and $S_2^i$ in a causal way, following $P_{S_1S_2|S}$ at each time step. 

Let $d(S,\hS)$ be the distortion function to measure the estimation error. In the same way as\cite{li2024analysis, ahmadipour2022coding}, we define the \ac{cd} region for the \ac{sddmmac} as $\calC(D)$ being the closure of all the rate tuples $(R_0,R_1,R_2)$ such that the messages $(M_0, M_1, M_2)$ can be reliably decoded while $\limsup_{n\to \infty}\frac{1}{n}\sum_{i=1}^n\expcs{}{d(S_i,\hS_i)} \le D$.


%% file: sections/figures/channel_model.tikz
\begin{tikzpicture}[node distance=3em and 2em]
    \def\msgsps{3em}
      \node[coordinate] (fix) {};
      \node[coordinate, left=of fix] (in0) {};
      \node[coordinate, above=of in0] (in1) {};
      \node[coordinate, below=of in0] (in2) {};
      \node[block, right=of in1] (enc1) {Enc. 1};
      \node[block, right=of in2] (enc2) {Enc. 2};
      \node[coordinate, below left=3em and 1.5em of enc1.center] (fixr) {};
      \node[smallblock, right=1em of fixr](state){$P_{SS_1S_2}$};
      \node[txtbigblock, right=7em of fix] (channel) {$P_{Y_1Y_2Z|X_1X_2S}$ \\ \\ $Z=(Y,S_R)$};
      \node[coordinate, below=2.5em of channel](chan_u){};
      \node[coordinate, above=2.5em of channel](chan_a){};
      \node[block, minimum height=3em, minimum width=2.5em, right=of channel](dec) {Dec.};
      \node[coordinate, right=4em of dec](out) {};

      \draw[-] (in0) -- node[above] {$M_0$} (fix);
      \draw[-] (fix) -- (fixr);
      \draw[->] (fixr) -- node[above] {} ([xshift=-1.5em]enc1.south);
      \draw[->] (fixr) -- node[above] {} ([xshift=-1.5em]enc2.north);
      \draw[->] (in1) -- node[above] {$M_1$} (enc1);
      \draw[->] (in2) -- node[above] {$M_2$} (enc2);

      \draw[->] (enc1.east) -- node[above] {$X_{1,i}$} ([yshift=\msgsps]channel.west);
      \draw[->] (enc2.east) -- node[above] {$X_{2,i}$} ([yshift=-\msgsps]channel.west);

      \draw[->](state)-- node[above]{$S_i$}(channel);
      \draw[->](state)-- node[left]{$S_1^i$}([xshift=1.35em]enc1.south);
      \draw[->](state)-- node[left]{$S_2^i$}([xshift=1.35em]enc2.north);

      \draw[-](channel.north) -- (chan_a);
      \draw[->](chan_a) -| node[anchor=north west] {$Y_1^{i-1}$} (enc1.north);
      
      \draw[-](channel.south) -- (chan_u);
      \draw[->](chan_u) -| node[anchor=south west] {$Y_2^{i-1}$} (enc2.south);
      
      \draw[->](channel.east) --node[above]{$Z_i$} (dec.west);
      \draw[->](dec.east) --node[above]{$\hat{M}_1, \hat{M}_2$} node[below]{$\hat{M}_0, \hS^n$} (out.west);
    \end{tikzpicture}

%% file: sections/results.tex
Let $\calP_D$ be the set of all random variables and encoding functions $(\Omega, f_1, f_2)$ with $\Omega = (U, W_1,\allowbreak W_2, U_1, \allowbreak U_2, T_1, T_2, V_1,V_2)$, $f_q$ mapping $(U,W_q,U_q,S_q)$ to $X_q$ for $q\in \{1,2\}$ such that
the overall joint distribution of $(S, S_1,S_2, \Omega, X_1,X_2,Z,Y_1,Y_2,\hS)$ is factorized as
\begin{equation}\label{eq:joint_pdf}
\begin{split}
    &P_S(s) P_{S_1S_2 |S}(s_1,s_2|s) P_U(u) P_{W_1|U}(w_1 | u) P_{W_2|U}(w_2 | u) \\
    &\cdot P_{U_1|UW_1}(u_1 | u,w_1) P_{U_2|UW_2}(u_2 | u,w_2)\\
    &\cdot \indf{x_1 = f_1(u, w_1, u_1, s_1)} \indf{x_2 = f_2(u, w_2, u_2, s_2)}\\
    &\cdot P_{Y_1Y_2Z|X_1X_2S}(y_1,y_2,z|x_1,x_2,s) \\
    &\cdot P_{T_1 | S_1Y_1} (t_1 | s_1, y_1) P_{T_2 | S_2Y_2} (t_2 | s_2, y_2) \\
    &\cdot P_{V_1 | S_1 U W_1 W_2 U_1 Y_1 T_1} (v_1 | s_1, u, w_1, w_2, u_1, y_1, t_1)\\
    &\cdot P_{V_2 | S_2 U W_1 W_2 U_2 Y_2 T_2} (v_2 | s_2, u, w_1, w_2, u_2, y_2, t_2)\\
    &\cdot \indf{\hat{s}=h^*(\omega, z)},
\end{split} 
\end{equation}
and the constraint
\begin{equation}
    \expcs{}{d(S, h^*(\Omega, Z))} \le D
\end{equation}
is fulfilled,
where $\indf{\cdot}$ is the indicator function, $h^*$ is the optimal estimation function at the decoder defined in \eqref{eq:opt_est} by letting $W = (\Omega,Z)$.


Let $\calR(\calP_D)$ be the set of all rate tuples $(R_0, R_1, R_2)$ satisfying

\begin{subequations}\label{eq:rate_ineq}
\begin{align}
    R_0 &\ge 0, R_1\ge 0, R_2 \ge 0,\\
    R_1 &= R_1'+R_1'', R_2 = R_2' + R_2''\\
    R_0 + R_1' + R_2' &\le I(U; T_1,T_2,Z)\\
    R_1' &\le I(W_1; S_2, Y_2 | U,W_2,U_2)\\
    R_2' &\le I(W_2; S_1, Y_1 | U,W_1, U_1)\\
    R_1' &\le I(W_1; T_1, T_2, Z| U,W_2)\\
    R_2' &\le I(W_2; T_1, T_2, Z| U,W_1)\\
    R_1' + R_2' &\le I(W_1, W_2; T_1, T_2, Z| U)\\
    R_1'' &\le I(U_1; T_1, T_2, Z| U,W_1,W_2,U_2)\notag\\
    &\qquad - (R_{s1} + \tR_{s1}) \\
    R_2'' &\le I(U_2; T_1, T_2, Z| U,W_1,W_2,U_1)\notag\\
    &\qquad - (R_{s2} + \tR_{s2})\\
    R_1'' + R_2'' &\le I(U_1, U_2; T_1, T_2, Z | U,W_1,W_2)  \notag\\
    &\qquad - (R_{s1} + \tR_{s1} + R_{s2} + \tR_{s2})\\
    R_{s1} &> I(T_1; S_1,Y_1 | T_2,Z)\\
    R_{s2} &> I(T_2; S_2,Y_2 | T_1,Z)\\ 
    R_{s1} + R_{s2} &> I(T_1, T_2; S_1, Y_1, S_2,Y_2 | Z)\\
    \tR_{s1} &> \notag\\
    &\hspace{-8mm} I(V_1; S_1,Y_1 | U,W_1,W_2,U_1,U_2,T_1,T_2,V_2,Z)\\
    \tR_{s2} &> \notag\\
    &\hspace{-8mm} I(V_2; S_2,Y_2 | U,W_1,W_2,U_1,U_2,T_1,T_2,V_1,Z)\\
    \tR_{s1} + \tR_{s2} &> \notag\\
    &\hspace{-12mm} I(V_1,V_2; S_1,Y_1,S_2,Y_2 | U,W_1,W_2,U_1,U_2,T_1,T_2,Z)
\end{align}
\end{subequations}
for all $(U, W_1, W_2, U_1, U_2, T_1, T_2, V_1,V_2, f_1, f_2)\in \calP_D$.

\begin{theorem}
$\calR(\calP_D) $ is an achievable \ac{rd} region, i.e.,
\begin{equation}
    \calR(\calP_D) \subseteq \calC(D).
\end{equation}
\end{theorem}
The proof of achievability combines the coding schemes from \cite{lapidoth2012multiple, willems1982informationtheoretical, carleial1982multiple, li2024analysis, kramer2008topics}, and is elaborated in Section~\ref{sec:proof}.
In particular, the private message $M_q$ at encoder $q \in \{1,2\}$ is first split into two parts $M_q = (M_q', M_q'')$ with $M_q' \in [2^{nR_q'}]$ and $M_q'' \in [2^{nR_q''}]$, where $M_q'$ represents the message that can be decoded by the other encoder through the feedback channel and $M_q''$ is the remaining private message that is unknown to the other. This leads to a total common message rate of $R_0 + R_1' + R_2'$ encoded in the auxiliary random variables $U, W_1, W_2$, while $M_q''$ is encoded in $U_q$. The transmit signal $X_q$ is chosen as a function of $(U, W_q, U_q, S_q)$ following the Shannon strategy\cite{el2011network}. Due to the presence of \ac{sit}, splitting rates for encoding \ac{sit} at two encoders, represented by $T_1$ and $T_2$, can improve the achievable rate region\cite{lapidoth2012multiple, sen2013channel}, which have the rates of $R_{s1}$ and $R_{s2}$ in our region. Finally, additional state descriptions $V_1$ and $V_2$ with rates of $\tR_{s1}$ and $\tR_{s2}$ are needed to enhance the state estimation task at the decoder\cite{bross2017rate,li2024analysis}. The difference between these two types of state descriptions is that $T_1$ and $T_2$ are used primarily for improving the communication performance, while $V_1$ and $V_2$ are purely adopted for the state estimation task and are decoded after message decoding, which can be removed if no state estimation task takes place.

\begin{remark}
    For the case of strictly causal \ac{sit}, i.e., $S_1^{i-1}$ and $S_2^{i-1}$ are present at time step $i$, the achievable region can be simply derived from $\calR(\calP_D)$ by removing the dependence of $(X_1, X_2)$ on $(S_1, S_2)$.
\end{remark}

%% file: sections/app.tex
\subsection{Monostatic-uplink ISAC}

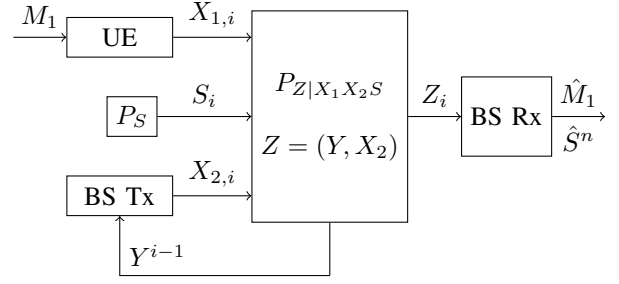
\begin{figure}[t]
    \centering
    \input{sections/figures/monoup.tikz}
    \caption{The equivalent model of monostatic-uplink \ac{isac}.}
    \label{fig:isac}
\end{figure}

A simple monostatic-uplink \ac{isac} system consists of a \ac{bs} and a \ac{ue}, where the \ac{bs} wishes to detect and estimate parameters of a target through echo signals while receiving uplink signals carrying messages from the \ac{ue}. In fact, such a system can be modeled by the \ac{sddmmac} in Fig.~\ref{fig:channelmodel} by treating the \ac{bs} as a transmitter and a receiver, ``virtually" separated from each other\cite{li2024analysis}, as illustrated in Fig.~\ref{fig:isac}. 
To model the monostatic sensing link, the target parameters are considered as the channel state $S$ that is to be estimated at the \ac{bs} receiver, and the echos are modeled by the perfect feedback signal, which is utilized to design sensing signal $X_2$. Furthermore, $X_2$ is known to the \ac{bs} receiver and is thus treated as the \ac{sir}, or equivalently one of the channel outputs.

The system performance is assessed by the relationship between the uplink channel capacity and the minimum achievable estimation distortion at \ac{bs} receiver. A possible achievable \ac{rd} region can be derived from \eqref{eq:rate_ineq} by first observing that $R_0=R_2=0$, $S_1 = S_2 = Y_1 = \varnothing$, $Y_2 = Y$ and $S_R = X_2$. 
Noting that no \ac{sit} and feedback is present at \ac{ue} and $Y$ is a deterministic function of $Z$, the lower bounds of $R_{s1}$, $R_{s2}$, $\tR_{s1}$ and $\tR_{s2} $ become zero. Hence, they are set to zero to provide a higher upper bound for $R_1''$. 
We then obtain the region as
\begin{equation}
\begin{split}
    R_1' &\le I(U; T_1,T_2,X_2,Y)\\
    R_1' &\le I(W_1; Y| U,W_2,U_2)\\
    R_1' &\le I(W_1; T_1, T_2, X_2,Y | U, W_2)\\
    R_1'' &\le I(U_1; T_1, T_2, X_2, Y | U, W_1, W_2, U_2)\\
    &\hspace{-1.2em}\expcs{}{d(S, h^*(\Omega, X_2, Y))} \le D.
\end{split}
\end{equation}
Due to the Markov chains $(T_1,T_2) - Y - (U,W_1,W_2,\allowbreak U_1,U_2,\allowbreak X_1, X_2, S)$ and $(V_1, V_2) - (U,W_1,\allowbreak W_2, U_1,U_2,\allowbreak X_2,Y) - S$, we can remove $(T_1,T_2,V_1,V_2)$ from the rate region as well as the dependence of $h^*$\cite[Lemma 6]{li2024analysis}. Furthermore, since $X_2$ is a deterministic function of $(U,W_2,U_2)$ and $W_1 - U - (W_2,U_2)$ forms a Markov chain, it shows that $I(W_1;Y|U,W_2,U_2) \ge I(W_1; X_2,Y | U, W_2)$. The \ac{rd} region can be rewritten as
\begin{equation}
\begin{split}
    R_1' &\le I(U; X_2,Y)\\
    R_1' &\le I(W_1; X_2,Y | U, W_2)\\
    R_1'' &\le I(U_1; X_2, Y | U, W_1, W_2, U_2)\\
    &\hspace{-1.2em}\expcs{}{d(S, h^*(U,W_1,W_2,U_1,U_2, X_2, Y))} < D,
\end{split}
\end{equation}
which can be further simplified by setting $U=W_1=W_2 = \varnothing$ and $U_1 = X_1$, $U_2=X_2$, resulting in
\begin{equation}\label{eq:monoup_cdf}
\begin{split}
    &R_1 = R_1' + R_2'' \le I(X_1 ; Y | X_2)\\
    &\expcs{}{d(S, h^*(X_1, X_2, Y))} \le D.
\end{split}
\end{equation}
In fact, \eqref{eq:monoup_cdf} is the \ac{cd} region for the model in Fig.~\ref{fig:isac}. The proof of converse follows the similar procedure as \cite{ahmadipour2022information, li2024analysis} by identifying the concavity of $R_1$ in $D$ and the Markov chain $S_i - (X_{1,i}, X_{2,i}, Y_i) - (X_2^{n\backslash i}, Y^{n\backslash i})$ at each time step $i$, which is thus omitted here due to the page limitation.

\subsection{Multi-sensor Network}

Another application is motivated by distributed multi-modal networks\cite{nguyen2023non}, in which wireless devices are equipped with different types of sensors to acquire environmental information. The multi-modal sensing data are fused at \acp{ap} for advanced tasks. One example is that multiple \acp{ap} carrying cameras communicate with a warehouse robot and help it with localizing itself. \acp{ap} get rough estimates of the robot position by cameras at each time step, which are then encoded and transmitted to the robot with other communication data. In this case, the channel state $S$ is the true position of the robot, and the \ac{sit} is the noisy estimate of $S$ obtained by other sensors and available causally. Feedback is assumed absent, so $(U,W_1,W_2,U_1,U_2)$ are independent of $(T_1, T_2, S_1, S_2)$. Consequently, the achievable \ac{rd} region is given in~\eqref{eq:cd_region_multisensor}, derived from $\calR(\calP_D)$ by setting $W_1 = W_2 = \varnothing$.
\begin{equation}\label{eq:cd_region_multisensor}
\begin{split}
    & R_0 \le I(U; Z | T_1,T_2)\\
    & R_1 \le I(U_1; Z | U, U_1, T_1, T_2) - (R_{s1} + \tR_{s1})\\
    & R_2 \le I(U_2; Z | U, U_2, T_1, T_2) - (R_{s2} + \tR_{s2})\\
    & R_1 + R_2 \le I(U_1, U_2; Z | U, T_1, T_2) -\\
    &\hspace{8em}(R_{s1} + \tR_{s1} + R_{s2} + \tR_{s2})\\
    & R_{s1} > I(T_1; S_1 | T_2, Z)\\
    & R_{s2} > I(T_2; S_2 | T_1, Z)\\ 
    & R_{s1} + R_{s2} > I(T_1,T_2; S_2 | Z)\\ 
    & \tR_{s1} > I(V_1; S_1 | U, U_1, U_2, T_1, T_2, V_2, Z)\\ 
    & \tR_{s2} > I(V_2; S_2 | U, U_1, U_2, T_1, T_2, V_1, Z)\\ 
    & \tR_{s1} + \tR_{s2} > I(V_1, V_2; S_2 | U, U_1, U_2, T_1, T_2, Z)\\
    & \expcs{}{d(S, h^*(U,U_1,U_2,T_1,T_2,V_1,V_2,Z))}\le D.
\end{split}
\end{equation}

%% file: sections/figures/monoup.tikz
\begin{tikzpicture}[node distance=3em and 2em]
    \def\msgsps{3em}
      \node[coordinate] (fix) {};
      \node[coordinate, left=of fix] (in0) {};
      \node[coordinate, above=of in0] (in1) {};
      \node[coordinate, below=of in0] (in2) {};
      \node[block, right=of in1] (enc1) {UE};
      \node[block, right=of in2] (enc2) {BS Tx};
      \node[coordinate, below left=3em and 1.5em of enc1.center] (fixr) {};
      \node[smallblock, right=1em of fixr](state){$P_S$};
      \node[txtbigblock, right=7em of fix] (channel) {$P_{Z|X_1X_2S}$ \\ \\ $Z=(Y,X_2)$};
      \node[coordinate, below=2em of channel](chan_u){};
      \node[block, minimum height=3em, minimum width=2.5em, right=of channel](dec) {BS Rx};
      \node[coordinate, right=of dec](out) {};

      \draw[->] (in1) -- node[above] {$M_1$} (enc1);

      \draw[->] (enc1.east) -- node[above] {$X_{1,i}$} ([yshift=\msgsps]channel.west);
      \draw[->] (enc2.east) -- node[above] {$X_{2,i}$} ([yshift=-\msgsps]channel.west);

      \draw[->](state)-- node[above]{$S_i$}(channel);
      
      \draw[-](channel.south) -- (chan_u);
      \draw[->](chan_u) -| node[anchor=south west] {$Y^{i-1}$} (enc2.south);
      
      \draw[->](channel.east) --node[above]{$Z_i$} (dec.west);
      \draw[->](dec.east) --node[above]{$\hat{M}_1$} node[below]{$\hS^n$} (out.west);
    \end{tikzpicture}

%% file: sections/proof.tex

\begin{lemma}\label{lemma:proof_lemma1}
    Let $(R_0, R_1, R_2) \in \calR(\calP_D)$, then $R_1 + R_2$ can not exceed $I(W_1, \allowbreak W_2, U_1,U_2; Z | U)$ and $R_0+R_1 +R_2$ can not exceed $I(U,W_1,\allowbreak W_1,U_1,U_2; Z)$. Consequently, if $I(W_1, W_2, \allowbreak U_1,U_2; Z | U) = 0$, implying $R_1 + R_2 = 0$, then $R_0 \le I(U;Z)$ is achievable by following the same coding scheme as the single-user channel.
\end{lemma}
\begin{lemma}\label{lemma:proof_lemma2}
    Let $(R_0, R_1, R_2) \in \calR(\calP_D)$. If $I(W_2,U_2; \allowbreak Z,S_1,\allowbreak Y_1 |\allowbreak U,\allowbreak W_1,U_1) = 0$, then $R_2$ must be zero and $R_0 + R_1$ can not exceed $I(U,W_1,U_1; Z)$, which is achievable by treating the \ac{mac} as a single-user channel with input $X_1$ and output $Z$.
\end{lemma}
The proof of these two lemmas can be conducted analogously to\cite{lapidoth2012multiple} and thus are omitted here. They guarantee the existence of finite codewords to encode and convey the first state description reliably at the last blocks. Lemma~\ref{lemma:proof_lemma1} implies that $I(W_1, U_1; Z | U, W_2, U_2)$ and $I(W_2, U_2; Z | U, W_1, U_1)$ can not be both zero when $R_1 + R_2 > 0$, otherwise the proof for $R_0 \le I(U;Z)$ suffices. Without loss of generality, we set $I(W_1, U_1; Z | U, W_2, U_2) > \alpha_1 > 0$ for a strictly positive value $\alpha_1$ in the following. Similarly, we set $I(W_2,U_2; Z,S_1,Y_1| U,W_1,U_1) > \alpha_2 > 0$, and if it is violated, then the proof for $R_0 + R_1 \le I(U,W_1,U_1; Z)$ is sufficient according to Lemma~\ref{lemma:proof_lemma2}.
We fix the set $\calP_D$ achieving the region $\calR(\calP_D)$.
The transmission happens in $B+4$ blocks. 

\subsection{Codebook Generation}
For each block $b\in [B + 1]$ and each encoder $ q\in \{1,2\}$:
\begin{itemize}
    \item Generate $2^{n(R_0 + R_1' + R_2')}$ codewords $u^n ( m_{c,b} )$ \ac{iid} from $P_{U}$ with $m_{c,b} = (m_{0,b}, m'_{1,b-1}, m'_{2,b-1} )$, $m_{0,b} \in [2^{nR_0}]$, $m_{q,b-1}' \in [2^{nR_q'}]$.
    \item For each $u^n ( m_{c,b} )$, generate $2^{nR_q'}$ codewords $w_q^n(m_{q,b}' | m_{c,b})$ \ac{iid} from $P_{W_q|U}$ with $m_{q,b}' \in [2^{nR_q'}]$.
    \item Generate $2^{n(R_{sq} + R_{sq}')}$ codewords $t_q^n(k_{q,b}, l_{q,b})$ \ac{iid} from $P_{T_q}$ with $k_{q,b}\in [2^{nR_{sq}}]$, $l_{q,b}\in [2^{nR_{sq}'}]$.
    \item For each $( m_{c,b}, m_{q,b}')$, generate $2^{n(R_q''+R_{sq} + \tR_{sq})}$ codewords $u_q^n(m_{q,b}'', k_{q,b-1}, j_{q,b-1} | m_{c,b}, m_{q,b}')$ \ac{iid} from $P_{U_q | W_q U}$ with $m_{q,b}'' \in [2^{nR_q''}]$, $j_{q,b-1} \in [2^{n\tR_{sq}}]$.
    \item For each $\mu_{q,b} = ( m_{c,b}, m_{1,b}',  m_{2,b}', m_{q,b}'', k_{q,b-1}, j_{q,b-1}, \allowbreak k_{q,b}, j_{q,b})$, generate $2^{n(\tR_{qs} + \tR_{qs}')}$ codewords $v_q^n(j_{q,b}, o_{q,b} | \mu_{q,b}) $ \ac{iid} from $P_{V_q | UW_1W_2U_qT_q}$, with $j_{q,b} \in [2^{n\tR_{sq}}]$ and $o_{q.b} \in [2^{n\tR_{sq}'}]$.
\end{itemize} 
For block $B+2$, let $n_1 = nR_{s1}/\alpha_1$:
\begin{itemize}
    \item Generate one length-$n_1$ codeword $u_{B+2}^{n_1}$ \ac{iid} from $P_{U}$, one length-$n_1$ codeword $w_{2,B+2}^{n_1}$ \ac{iid} from $P_{W_2|U}(\cdot | u_{B+2, i})$, and one length-$n_1$ codeword $u_{2,B+2}^{n_1}$ \ac{iid} from $P_{U_2|W_2U}(\cdot|w_{2, B+2, i}, u_{B+2, i})$.
    \item Generate $2^{nR_{s1}}$ length-$n_1$ codewords $u_1^{n_1}(k_{1,B+1})$ \ac{iid} from $P_{U_1W_1|U}(\cdot|u_{B+2, i})$.
\end{itemize}
For block $B+3$, let $n_2 = nR_{s2}/\alpha_2$:
\begin{itemize}
    \item Generate one length-$n_2$ codeword $u_{B+3}^{n_2}$ \ac{iid} from $P_{U}$, one length-$n_2$ codeword $w_{1,B+3}^{n_2}$ \ac{iid} from $P_{W_1|U}(\cdot | u_{B+3, i})$., and one length-$n_2$ codeword $u_{1, B+3}^{n_2}$ \ac{iid} from $P_{U_1|W_1U}(\cdot | w_{1,B+3,i},u_{B+3,i})$.
    \item Generate $2^{nR_{s2}}$ length-$n_2$ codewords $u_2^{n_2}(k_{2,B+1})$ \ac{iid} from $P_{U_2W_2|U}(\cdot|u_{B+3,i})$.
\end{itemize}
For block $B+4$, let $n_3 = n_2(H(S_1Y_1) + \delta)/\alpha_1$ with $\delta > 0$:
\begin{itemize}
    \item Generate one length-$n_3$ codeword $u_{B+4}^{n_3}$ \ac{iid} from $P_{U}$, one length-$n_3$ codeword $w_{2,B+4}^{n_3}$ \ac{iid} from $P_{W_2|U}(\cdot | u_{B+4, i})$, and one length-$n_3$ codeword $u_{2,B+4}^{n_3}$ \ac{iid} from $P_{U_2|W_2U}(\cdot|w_{2,B+4,i},u_{B+4,i})$.
    \item Generate $2^{n_2(H(S_1Y_1) + \delta)}$ length-$n_3$ codewords $u_1^{n_3}(k_{1, B+3})$ \ac{iid} from $P_{U_1W_1|U}(\cdot|u_{B+4, i})$ for $k_{1, B+3} \in [2^{n_2(H(S_1Y_1) + \delta)}]$.
\end{itemize}

\subsection{Encoding}

For each block $b\in [B + 1]$, encoder $q\in \{1,2\}$ splits the message $m_{q,b} = (m_{q,b}', m_{q,b}'')$ and transmits
\begin{equation}
    x_{q,b,i} = f_q(u_{b,i}, w_{q,b,i},  u_{q,b,i}, s_{q,b,i}),
\end{equation}
with $u_{b,i} = u_i(m_{c,b})$, $w_{q,b,i} = w_{q,i}(m_{q,b}' | m_{c,b})$, $u_{q,b,i} = u_{q,i}( m_{q,b}'' , k_{q,b-1}, j_{q,b-1}, | m_{c,b}, m_{q,b}')$.
The associated indices are obtained as follows.

$m_{q,0}' =m_{q,B+1}' = k_{q,0} = j_{q,0} = 1$.


At the end of block $b\in [B+1]$, with the knowledge of $(s_{q,b}^n, y_{q,b}^n)$,
the encoder $q$ first decodes $m_{q',b}'$ with $q' \neq q$ transmitted by the other encoder through the feedback link by finding the unique $\hat{m}_{q',b}'$ such that
\begin{equation}\label{eq:m'q_decode}
\begin{split}
    &(u^n(m_{c,b}), w_q^n(m_{q,b}' | m_{c,b} ) ,w_{q'}^n(\hat{m}_{q',b}' |  m_{c,b}) , \\
    &\quad u_q^n(m_{q,b}'', k_{q,b-1}, j_{q,b-1} | m_{c,b}, m_{q,b}'), s_{q,b}^n, y_{q,b}^n) \\
    &\hspace{10em}\in \typset(P_{UW_1W_2U_qS_qY_q}).
\end{split}
\end{equation}
Then, it looks for the first state description $(k_{q,b}, l_{q,b})$ with
\begin{equation}\label{eq:first_state_description}
    (t^n_q(k_{q,b}, l_{q,b}), s_{q,b}^n, y_{q,b}^n) \in \typset(P_{T_qS_qY_q})
\end{equation}
and the second description $(j_{q,b}, o_{q,b})$ with
\begin{equation}\label{eq:second_state_description}
    (v_q^n(j_{q,b}, o_{q,b} | \mu_{q,b}), s_{q,b}^n, y_{q,b}^n) \in \typset(P_{V_qS_qY_q}).
\end{equation}



At block $B+2$, encoder 1 transmits nothing but the first state description $k_{1,B+1}$, i.e.,
\begin{equation}
    x_{1,B+2,i} = f_1(u_{B+2,i}, u_{1,i}(k_{1,B+1}), s_{1,B+2,i}),
\end{equation}
and encoder 2 transmits the deterministic signals
\begin{equation}
    x_{2,B+2,i} = f_2(u_{B+2,i}, w_{2,B+2,i}, u_{2,B+2,i}, s_{2,B+2,i}).
\end{equation}

At block $B+3$, encoders transmit
\begin{equation}
\begin{split}
    x_{1,B+3,i} &= f_1(u_{B+3,i}, w_{1,B+3,i}, u_{1,B+3,i}, s_{1,B+3,i}),\\
    x_{2,B+3,i} &= f_2(u_{B+3,i}, u_{2,i}(k_{2,B+1}), s_{2,B+3,i}).
\end{split}
\end{equation}

At block $B+4$, with the knowledge of $(s^{n_2}_{1,B+3}, y^{n_2}_{1,B+3})$, encoder 1 compresses it losslessly into the index $k_{1,B+3}$, and 
\begin{equation}
\begin{split}
    x_{1,B+4,i} &= f_1(u_{B+4,i},u_{1,i}(k_{1,B+3}), s_{1,B+4,i}),\\
    x_{2,B+4,i} &= f_2(u_{B+4,i},w_{1,B+4,i},u_{2,B+4,i}, s_{2,B+4,i})
\end{split}
\end{equation}
are then transmitted.

\subsection{Decoding}

The decoder first applies the backward decoding strategy to recover messages and the first state description at each block, and then decodes the second state description in the forward direction. Finally, based on the decoding results, it estimates the channel state at blocks $b\in [B]$.

At the end of block $B+4$, the decoder looks for a unique $\hat{k}_{1,B+3}$ such that
\begin{equation}
\begin{split}
    (u^{n_3}_{1,B+4}(\hat{k}_{1,B+3}), z^n_{B+4})\in \typset(P_{U_1Z}),\\[0.05in]
\end{split}
\end{equation}
from which the lossless version of $(\hat{s}_{1,B+3}^{n_2}, \hat{y}_{1,B+3}^{n_2})$ are recovered. Then at block $B+3$, the decoder finds the unique $\hat{k}_{2,B+1}$ such that
\begin{equation*}
    (u_{2,B+3}^{n_2}(\hat{k}_{2,B+1}), \hat{s}_{1,B+3}^{n_2}, \hat{y}_{1,B+3}^{n_2}, z^n_{B+3})\in \typset(P_{U_1S_1Y_1Z}).
\end{equation*}
Similarly, at block $B+2$, $\hat{k}_{1,B+1}$ is decoded when
\begin{equation}
    (u_1^{n_1}(\hat{k}_{1,B+1}), z^n_{B+2})\in \typset(P_{U_1Z}).
\end{equation}

Assuming $(\hat{k}_{1,B+1}, \hat{k}_{2,B+1}) = (k_{1,B+1}, k_{2,B+1})$ are decoded correctly, at the end of block $B+1$, it first recovers $(\hat{l}_{1,B+1}, \hat{l}_{2,B+1})$ by finding
\begin{equation}
\begin{split}
    &(t_1^n(k_{1,B+1}, \hat{l}_{1,B+1}) , t_2^n(k_{2,B+1}, \hat{l}_{2,B+1}), z_{B+1}^n)\\
    &\hspace{50mm} \in \typset(P_{T_1T_2Z}). 
\end{split}
\end{equation}
If $(\hat{l}_{1,B+1}, \hat{l}_{2,B+1}) = (l_{1,B+1}, l_{2,B+1})$ are also decoded correctly, the decoder searches for the unique $(\hat{m}_{1,B+1}'', \hat{k}_{1,B}, \hat{j}_{1,B}, \hat{m}_{2,B+1}'', \hat{k}_{2,B}, \hat{j}_{2,B}, \hat{m}_{c,B+1})$ such that
\begin{equation}\label{eq:typdec_B+1}
\begin{split}
    & (u^n(\hat{m}_{c,B+1}), w_1^n(1 |  \hat{m}_{c,B+1}), w_2^n(1 |  \hat{m}_{c,B+1}),\\
    &\quad u_1^n(\hat{m}_{1,B+1}'', \hat{k}_{1,B}, \hat{j}_{1,B} | \hat{m}_{c,B+1}, 1),\\
    &\quad u_2^n(\hat{m}_{2,B+1}'', \hat{k}_{2,B}, \hat{j}_{2,B} | \hat{m}_{c,B+1}, 1), \\
    &\quad t_1^n(k_{1,B+1}, l_{1,B+1}) , t_2^n(k_{2,B+1}, l_{2,B+1}), z_{B+1}^n)\\
    &\hspace{40mm} \in \typset(P_{UW_1W_2U_1U_2T_1T_2Z}).
\end{split}
\end{equation}

From block $B$ to $1$, the decoder performs the same procedures to first decode $(\hat{l}_{1,b}, \hat{l}_{2,b})$ and then $(\hat{m}_{1,b}'', \hat{k}_{1,b-1},\allowbreak \hat{j}_{1,b-1}, \hat{m}_{2,b}'', \hat{k}_{2,b-1}, \hat{j}_{2,b-1}, \hat{m}_{c,b-1})$. The only difference is that all ``1"s in \eqref{eq:typdec_B+1} are replaced by $\hat{m}_{1,b}'$ or $\hat{m}_{2,b}'$.

Subsequently, the decoder recovers $(\hat{o}_{1,b}, \hat{o}_{2,b})$ in the forward direction for $b\in [B]$. In particular, it looks for the unique typical sequence
\begin{equation*}
    (v_1^n(j_{1,b} , \hat{o}_{1,b} | \mu_{1,b}), v_2^n(j_{2,b} , \hat{o}_{2,b} | \mu_{2,b}), z^n_{b}) \in \typset(P_{V_1V_2Z}).
\end{equation*}

Finally, the estimated state sequences $\hat{s}_b^n$ at block $b\in [B]$ are obtained by
\begin{equation}
    \hat{s}_{b,i}=  h^*(\omega_{b,i}, z_{b,i}),
\end{equation}
with $\omega_{b,i} = (u_{b,i}, w_{1,b,i}, w_{2,b,i}, u_{1,b,i}, u_{2, b,i}, t_{1,b,i}, t_{2,b,i}, v_{1,b,i}, \allowbreak v_{2,b,i})$, where the indices of codewords are omitted for convenience.

\subsection{Analysis}

The first error event at encoder $q$ is incorrect decoding of $m_{q',b}$. From \eqref{eq:m'q_decode} it follows that this error probability tends to zero for $n\to \infty$ if
\begin{equation}\label{eq:rate_const_r1'}
\begin{split}
    R_1' < I(W_1; Y_2, S_2 | U, W_2, U_2),\\
    R_2' < I(W_2; Y_1, S_1 | U, W_1, U_1),
\end{split}
\end{equation}
due to the packing lemma\cite{el2011network}. Furthermore, the existence of $k_{q,b}$ and $j_{q,b}$ in \eqref{eq:first_state_description} and \eqref{eq:second_state_description} can be guaranteed if 
\begin{equation}
\begin{split}
    R_{s1} + R_{s1}' > I(T_1; S_1, Y_1)\\
    R_{s2} + R_{s2}' > I(T_2; S_2, Y_2)\\[0.06in]
\end{split}
\end{equation}
and
\begin{equation}
\begin{split}
    \tR_{s1} + \tR_{s1}' > I(V_1; S_1, Y_1 | U,W_1, W_2, U_1, T_1 )\\
    \tR_{s2} + \tR_{s2}' > I(V_2; S_2, Y_2 | U,W_1, W_2, U_2, T_2 )
\end{split}
\end{equation}
following the covering lemma\cite{el2011network}.

At the decoder, the error of finding the unique $(\hat{k}_{1,B+1}, \hat{k}_{2,B+1})$ tends to zero for $n\to \infty$ following the same analysis as in\cite{lapidoth2012multiple} due to Lemma~\ref{lemma:proof_lemma1}~and~\ref{lemma:proof_lemma2}.
For $(\hat{m}_{1,b}'', \hat{k}_{1,b-1},\allowbreak \hat{j}_{1,b-1}, \hat{m}_{2,b}'',\allowbreak \hat{k}_{2,b-1},\allowbreak \hat{j}_{2,b-1}, \hat{m}_{c,b-1})$ and $(\hat{l}_{1,b}, \hat{l}_{2,b})$ to be decoded correctly, it should hold that
\begin{equation}
\begin{split}
    R_0 + R_1' + R_2' &< I(U; T_1, T_2, Z)\\
    R_1' &< I(W_1; T_1,T_2,Z | U, W_2)\\
    R_2' &< I(W_2; T_1,T_2,Z | U, W_1)\\
    R_1' + R_2' &< I(W_1, W_2; T_1,T_2,Z | U)\\
    R_1'' + R_{s1} + \tR_{s1} &< I(U_1; T_1,T_2, Z |U, W_1, W_2, U_2)\\
    R_2'' + R_{s2} + \tR_{s2} &< I(U_2; T_1,T_2, Z |U, W_1, W_2, U_1)\\
    R_1'' + R_{s1} + \tR_{s1} &+ R_2'' + R_{s2} + \tR_{s2}\\
    &< I(U_1, U_2; T_1, T_2,Z| U, W_1, W_2)\\
\end{split}
\end{equation}
as well as
\begin{equation*}
\begin{split}
    R_{s1}' &< I(T_1; T_2, Z)\\
    R_{s2}' &< I(T_2; T_1, Z)\\
    R_{s1}' + R_{s2}' &<I(T_1; T_2,Z) + I(T_2; T_1,Z) - I(T_1; T_2|Z)
\end{split}
\end{equation*}
following similar arguments in the analysis of distributed Wyner-Ziv problem\cite{gastpar2004wyner}.

In the forward direction, the zero error of decoding $(\hat{o}_{1,b}, \hat{o}_{2,b})$ can be guaranteed when
\begin{equation}\label{eq:rate_const_trs1'}
\begin{split}
    \tR_{s1}' &< I(V_1; V_2, Z | U, W_1, W_2, U_1, U_2, T_1, T_2)\\
    \tR_{s2}' &< I(V_2; V_1, Z | U, W_1, W_2, U_1, U_2, T_1, T_2)\\
    \tR_{s1}' + \tR_{s2}' &< I(V_1; V_2, Z | U, W_1, W_2, U_1, U_2, T_1, T_2)\\
    &\quad + I(V_2; V_1, Z | U, W_1, W_2, U_1, U_2, T_1, T_2)\\
    &\quad - I(V_1; V_2 | U, W_1, W_2, U_1, U_2, T_1, T_2, Z).
\end{split}
\end{equation}

Combining the above inequalities and leveraging the Markov chains induced from \eqref{eq:joint_pdf}, we end up with the rate region in \eqref{eq:rate_ineq},
and it can be shown that as $n\to \infty$ and $B \to \infty$, the message decoding error probability tends to zero, and at the same time the expected distortion is constrained under $D$. This concludes the proof of achievability. 



%% file: sections/conclusion.tex
This paper establishes an achievable \ac{rd} region of joint state estimation and message decoding over \ac{sddmmac} in the presence of \ac{sit} and generalized feedback. The coding scheme explores the possibility of encoder collaboration and performance improvement via two-stage state descriptions. Two examples from \ac{isac} applications are then studied. In future work, it is necessary to explore \ac{isac} for multiuser scenarios under security constraints\cite{gunlu2023secure,chen2024distribution} and estimation guarantees \cite{chen2024distribution}. Only in this way can the strict requirements for 6G applications\cite{fettweis20216g} and the trustworthiness of 6G\cite{fettweis20216g,fettweis20226g} be achieved.

%% file: main.bbl
\begin{thebibliography}{10}
\providecommand{\url}[1]{#1}
\csname url@samestyle\endcsname
\providecommand{\newblock}{\relax}
\providecommand{\bibinfo}[2]{#2}
\providecommand{\BIBentrySTDinterwordspacing}{\spaceskip=0pt\relax}
\providecommand{\BIBentryALTinterwordstretchfactor}{4}
\providecommand{\BIBentryALTinterwordspacing}{\spaceskip=\fontdimen2\font plus
\BIBentryALTinterwordstretchfactor\fontdimen3\font minus
  \fontdimen4\font\relax}
\providecommand{\BIBforeignlanguage}[2]{{%
\expandafter\ifx\csname l@#1\endcsname\relax
\typeout{** WARNING: IEEEtran.bst: No hyphenation pattern has been}%
\typeout{** loaded for the language `#1'. Using the pattern for}%
\typeout{** the default language instead.}%
\else
\language=\csname l@#1\endcsname
\fi
#2}}
\providecommand{\BIBdecl}{\relax}
\BIBdecl

\bibitem{sutivong2003channel}
A.~Sutivong, \emph{Channel capacity and state estimation for state-dependent
  channels}.\hskip 1em plus 0.5em minus 0.4em\relax Stanford University, 2003.

\bibitem{kim2008state}
Y.-H. Kim, A.~Sutivong, and T.~M. Cover, ``State amplification,'' \emph{IEEE
  Transactions on Information Theory}, vol.~54, no.~5, pp. 1850--1859, 2008.

\bibitem{zhang2011joint}
W.~Zhang, S.~Vedantam, and U.~Mitra, ``Joint transmission and state estimation:
  A constrained channel coding approach,'' \emph{IEEE Transactions on
  Information Theory}, vol.~57, no.~10, pp. 7084--7095, 2011.

\bibitem{choudhuri2013causal}
C.~Choudhuri, Y.-H. Kim, and U.~Mitra, ``Causal state communication,''
  \emph{IEEE Transactions on Information Theory}, vol.~59, no.~6, pp.
  3709--3719, 2013.

\bibitem{bross2017rate}
S.~I. Bross and A.~Lapidoth, ``The rate-and-state capacity with feedback,''
  \emph{IEEE Transactions on Information Theory}, vol.~64, no.~3, pp.
  1893--1918, 2017.

\bibitem{liu2022integrated}
F.~Liu, Y.~Cui, C.~Masouros, J.~Xu, T.~X. Han, Y.~C. Eldar, and S.~Buzzi,
  ``Integrated sensing and communications: Toward dual-functional wireless
  networks for 6g and beyond,'' \emph{IEEE Journal on Selected Areas in
  Communications}, vol.~40, no.~6, pp. 1728--1767, 2022.

\bibitem{schwenteck20236g}
P.~Schwenteck, G.~T. Nguyen, H.~Boche, W.~Kellerer, and F.~H. Fitzek, ``6g
  perspective of mobile network operators, manufacturers, and verticals,''
  \emph{IEEE Networking Letters}, vol.~5, no.~3, pp. 169--172, 2023.

\bibitem{ahmadipour2022information}
M.~Ahmadipour, M.~Kobayashi, M.~Wigger, and G.~Caire, ``An
  information-theoretic approach to joint sensing and communication,''
  \emph{IEEE Transactions on Information Theory}, 2022.

\bibitem{li2024analysis}
X.~Li, V.~C. Andrei, A.~Djuhera, U.~J. M{\"o}nich, and H.~Boche, ``An analysis
  of capacity-distortion trade-offs in memoryless {ISAC} systems,'' \emph{arXiv
  preprint arXiv:2402.17058}, 2024.

\bibitem{ahmadipour2022coding}
M.~Ahmadipour, M.~Wigger, and M.~Kobayashi, ``Coding for sensing: An improved
  scheme for integrated sensing and communication over {MAC}s,'' in \emph{2022
  IEEE International Symposium on Information Theory (ISIT)}.\hskip 1em plus
  0.5em minus 0.4em\relax IEEE, 2022, pp. 3025--3030.

\bibitem{willems1982informationtheoretical}
F.~M. Willems, ``Informationtheoretical results for the discrete memoryless
  multiple access channel,'' Ph.D. dissertation, Katholieke Universiteit
  Leuven, 1982.

\bibitem{carleial1982multiple}
A.~Carleial, ``Multiple-access channels with different generalized feedback
  signals,'' \emph{IEEE Transactions on Information Theory}, vol.~28, no.~6,
  pp. 841--850, 1982.

\bibitem{kramer2008topics}
G.~Kramer \emph{et~al.}, ``Topics in multi-user information theory,''
  \emph{Foundations and Trends{\textregistered} in Communications and
  Information Theory}, vol.~4, no. 4--5, pp. 265--444, 2008.

\bibitem{lapidoth2012multiple}
A.~Lapidoth and Y.~Steinberg, ``The multiple-access channel with causal side
  information: Double state,'' \emph{IEEE Transactions on Information Theory},
  vol.~59, no.~3, pp. 1379--1393, 2012.

\bibitem{el2011network}
A.~El~Gamal and Y.-H. Kim, \emph{Network information theory}.\hskip 1em plus
  0.5em minus 0.4em\relax Cambridge University Press, 2011.

\bibitem{sen2013channel}
N.~Sen, \emph{Channel Capacity in the Presence of Feedback and Side
  Information}.\hskip 1em plus 0.5em minus 0.4em\relax Queen's University
  (Canada), 2013.

\bibitem{nguyen2023non}
L.~N. Nguyen, P.~Susarla, A.~Mukherjee, M.~L. Ca{\~n}ellas, C.~{\'A}. Casado,
  X.~Wu, D.~B. Jayagopi, M.~B. L{\'o}pez \emph{et~al.}, ``Non-contact
  multimodal indoor human monitoring systems: A survey,'' \emph{arXiv preprint
  arXiv:2312.07601}, 2023.

\bibitem{gastpar2004wyner}
M.~Gastpar, ``The {Wyner-Ziv} problem with multiple sources,'' \emph{IEEE
  Transactions on Information Theory}, vol.~50, no.~11, pp. 2762--2768, 2004.

\bibitem{gunlu2023secure}
O.~G{\"u}nl{\"u}, M.~R. Bloch, R.~F. Schaefer, and A.~Yener, ``Secure
  integrated sensing and communication,'' \emph{IEEE Journal on Selected Areas
  in Information Theory}, vol.~4, pp. 40--53, 2023.

\bibitem{chen2024distribution}
Y.~Chen, T.~Oechtering, H.~Boche, M.~Skoglund, and Y.~Luo,
  ``Distribution-preserving integrated sensing and communication with secure
  reconstruction,'' \emph{arXiv preprint arXiv:2405.07275}, 2024.

\bibitem{fettweis20216g}
G.~P. Fettweis and H.~Boche, ``{6G}: The personal tactile internet—and open
  questions for information theory,'' \emph{IEEE BITS the Information Theory
  Magazine}, vol.~1, no.~1, pp. 71--82, 2021.

\bibitem{fettweis20226g}
------, ``On {6G} and trustworthiness,'' \emph{Communications of the ACM},
  vol.~65, no.~4, pp. 48--49, 2022.

\end{thebibliography}
